\documentclass[runningheads]{llncs}
\usepackage{graphicx}
\usepackage{longtable}
\usepackage{comment}
\usepackage{array}
\usepackage{float}
\usepackage{hyperref}
\usepackage{amssymb}

\begin{document}
\title{Matrix Factorization for Cache Optimization in Content Delivery Networks (CDN)}
%
%\titlerunning{Abbreviated paper title}
% If the paper title is too long for the running head, you can set
% Adolf Kamuzora, Wadie Skaf, Ermiyas Birhanu, Jiyan Salim Mahmud, Péter Kiss, Tamás Jursonovics, Peter Pogrzeba, Tomáš Horváth, Imre Lendák
% an abbreviated paper title here
%
\author{Adolf Kamuzora\inst{1} \and
Wadie Skaf\inst{1} \and
Ermiyas Birihanu\inst{1} \and Jiyan Salim Mahmud\inst{1} \and
Péter Kiss\inst{1} \and
Tamás Jursonovics\inst{2} \and Peter Pogrzeba\inst{2} \and
Imre Lendák\inst{1} \and
Tomáš Horváth\inst{1}}
\authorrunning{A. Kamuzora et al.}
% First names are abbreviated in the running head. 
% If there are more than two authors, 'et al.' is used.
%
\institute{Telekom Innovation Laboratories, Data Science and Engineering Department,
Faculty of Informatics, Eötvös Loránd University
Pázmány Péter str. 1/A, 1117 Budapest, Hungary
\\
\email{\{adolfnfsp, skaf, ermiyasbirihanu, jiyan, axx6v4, lendak, tomas.horvath\}@inf.elte.hu }, \\ home page:
\texttt{http://t-labs.elte.hu/}
\and
\ Deutsche Telekom, Berlin, Germany} 

\maketitle              % typeset the header of the contribution

\begin{abstract}
%The abstract should briefly summarize the contents of the paper in
%15--250 words.
Content delivery networks (CDNs) are key components of high throughput, low latency services on the internet. CDN cache servers have limited storage and bandwidth and implement state-of-the-art cache admission and eviction algorithms to select the most popular and relevant content for the customers served. The aim of this study was to utilize state-of-the-art recommender system techniques for predicting ratings for cache content in CDN. Matrix factorization was used in predicting content popularity which is valuable information in content eviction and content admission algorithms run on CDN edge servers. A custom implemented matrix factorization class and MyMediaLite were utilized. The input CDN logs were received from a European telecommunication service provider. We built a matrix factorization model with that data and utilized grid search to tune its hyper-parameters. Experimental results indicate that there is promise about the proposed approaches and we showed that a low root mean square error value can be achieved on the real-life CDN log data.

\end{abstract}

\section{Introduction}

Streaming video services, like Netflix, Disney+, Amazon Prime, and others generate a significant part of Internet traffic today \cite{Cisco2020AnnualReport}. Content delivery networks (CDNs) are the key components making the delivery of such services with high throughput, low latency a reality. CDNs consist of origin servers and cache servers positioned closest to the customers, i.e., on the edge of the system. The cache servers have limited storage and bandwidth and implement state-of-the-art cache admission and cache eviction algorithms to select the most popular and relevant content for the customers served. In recommender systems, the goal is to suggest similar and popular items to users, given that similar users have requested or rated similar items. Thus, an interest is to predict or determine how the users would rate items that they have not seen. Algorithms such as matrix factorization (MF) could aid this endeavor. The intuition behind matrix factorization is that, with a discovered set of latent features for both users and items, it is possible to predict ratings and thus determine popular and relevant content for the customers.

Our research aims to utilize state-of-the-art recommender system techniques for predicting ratings for cache content in CDNs and their cost reduction potential in real-life content delivery scenarios. Our effort is driven by the needs of a leading European telecommunication service provider, who developed and built its managed CDN to ensure Internet Protocol television (IPTV) service delivery. We conducted experiments on an extensive dataset collected in a real-life CDN system. We show that matrix factorization models are viable in the CDN cache environment context.

Our paper is structured into five sections; introduction, related works, methodology section where, the dataset, processing, matrix factorization, evaluation and hyper-parameter tuning are presented. Then results of experiments are presented and finally, the conclusion.

\section{Related Works}

It is already well understood, how CDNs provide the required bandwidth, and thereby ensure reliable, high quality video services \cite{zolfaghari2020content}. The tremendous network growth faced by CDN operators forces them to continuously innovate and decrease the unit cost of Gigabyte delivery. This effort is supported by new hardware generations, state-of-the-art protocols \cite{Spiteri2020ABR}, new video encoding standards \cite{Cianci2014HDTV}, but mainly affected by the number of request cached on a CDN server; the cache hit rate (CHR). A small improvement in the CHR decreases the resource consumption, reduces the load in deeper caching layers and origin servers and ultimately saves costs. 

In spite of the increasing footprint of Edge cloud services, CDN is the most efficient tool for streaming multimedia contents \cite{zolfaghari2020content}\cite{Yu2017EdgeSurvey}. Adaptive Bitrate (ABR) streaming is a solution which aims to address the challenge for video streaming under unstable network conditions. For ABR streaming, each video is encoded into multiple bitrates, and the customer device selects a bitrate according to the network conditions. This ultimately aims at maximizing users' quality of experience (QoE) \cite{Gao2018ABR}. 
A CDN deploys many servers so that at least some lie in each user's network proximity. CDNs improve service quality and availability by replicating the digital content in those servers. Thus requests to objects (images, video, web content) are handled by nearby servers in case of a cache \textit{hit} (content present in cache). Otherwise, if the requested content is not available in the nearest CDN server, it has to be fetched from the origin server, causing slight delay on the customer side and increasing the load on the origin server. Different researchers claim that the advent of fog computing and machine learning would bring mitigation to some if not all the challenges faced by CDN operators, e.g., origin performance limitations and caching of content \cite{zolfaghari2020content}.

Caching performance improvement in terms of maximizing hit rates is one of the challenges faced in CDN. Authors in \cite{Brilli2016Performance} analyzed the performance of a deployed CDN caching system by deriving its performance metrics using real video on demand (VoD) traces derived from an internet service provider (ISP) CDN caching system. In their analysis, caching system logs that contain a full report of anonymized requests from each user were used, along with the result of the caching operation and time needed to fulfill a particular request. The results of the analyses showed that, first, QoE is optimal if the time to serve is bounded, not if it is the lowest possible one. Also, there is not a straight proportional law between time to serve and resource size. Second, in the case of cache effectiveness, if the cache is serving an optimal number of users: if the per-flow, per-user hit ratio, both normalized to the resource size, are almost identical, then the cache is well performing, and adding more users will not lead to further benefits. Third, it remains the question of cache optimization by cache admission and cache eviction algorithms \cite{Brilli2016Performance}.

According to authors in \cite{PESKA2019201}, the research area of recommender systems is rather new compared to other disciplines in computer science. However, rapid growth of electronic commerce (e-commerce), social networks and broadcasting applications triggered and facilitated development of recommender approaches and techniques (content-based and collaborative filtering).
Research studies on algorithms for collaborative filtering have proven to be effective in recommender systems on predicting popular items \cite{Elahi2016Coll_FilterSSurvey}. While user-based and item-based collaborative filtering methods are intuitive, matrix factorization (MF) techniques are effective since they aid to discover latent factors underlying interactions between users and items \cite{Takacs2008Matr_Fact}. This aspect facilitates to determine what content is frequently requested and thus, to be in the cache, whether during cache admission or cache eviction. Cache replacement concept is suggested by authors in \cite{Brilli2016Performance}.
In collaborative filtering, a popular model is k-nearest neighbour (kNN) \cite{Deshpande2004ItemBased}. However, this technique has not shown to scale well with increasing amount of data. Recently, matrix factorization has become popular in recommender systems, used for both implicit and explicit feedback \cite{PESKA2019201}. In earlier works on matrix factorization, singular value decomposition (SVD) was proposed to learn feature matrices \cite{Sarwar2002SVD}. But MF models learnt by SVD showed to be prone to overfitting. This lead to the proposition of regularized learning methods.  \cite{Rendle2008RegMF}. The most popular approach in collaborative filtering is the use of model-based techniques consisting of matrix factorization (e.g., plain matrix factorization, matrix factorization with biases). These methods combine good scalability, predictive accuracy, and offer much flexibility for modeling various real-life situations \cite{PESKA2019201}. Stochastic gradient descent-based optimization techniques have been efficient to find parameters of the model, i.e., the latent factors of users and items, minimizing the squared error or similar loss functions \cite{Koren2009MFTech}. For our work, we proposed using matrix factorization techniques to predict popular content, potentially optimizing cache admission/eviction in content delivery networks.

%weighted regularized MF, bayesian personalized ranking MF, non-negative MF

\section{Methodology}

\subsection{The Dataset}

The analysis was carried out on real-life CDN cache logs collected on multiple hosts in a large European telecommunication service provider. The logs contain detailed, but anonymized requests issued by CDN users, along with the result of the caching operations (hit/miss) and time needed to serve a request. Table \ref{tab1} provides a brief description of the attributes for the logs.

\begin{table}[h]
\centering
\begin{tabular}{ |m{3cm}|m{10cm}| }
 \hline
% \multicolumn{2}{CDN log line features} \\
% \hline
 
Feature & Description \\
\hline

statuscode & HTTP response status codes\\
\hline
contenttype & Indicates the media type of the resource\\
\hline
protocol & Http version \\
\hline
contentlength & The size of the resource, in decimal number of bytes\\
\hline
timefirstbyte & time from request processing until the first byte\\
\hline
timetoserv & time needed to process the request\\
\hline
osfamily & Client's device Operating System\\
\hline
sid & id uniq to a streaming session\\
\hline
cachecontrol & Directives for caching mechanisms in both requests and responses\\
\hline
uamajor & User-Agent'version, eg. browser's version\\
\hline
uafamily & User-Agent,usually client’s app \\
\hline
devicefamily & Type of the device \\
\hline
fragment & video or file fragment identifier \\
\hline
path & URL (address of request)\\
\hline
timestamp & Arrival time of the request\\
\hline
contentpackage & VoD asset identifier \\
\hline
coordinates & Long. and lat. of the client based on geoip lookup \\
\hline
livechannel & Live TV channel name \\
\hline
devicemodel & Client's device model \\
\hline
devicebrand & Client's device brand \\
\hline
host & Specifies the domain name of the server (for virtual hosting) \\
\hline
method & Http request method eg. Get,post \\
\hline
manifest & resource which the browser should cache for offline access \\
\hline
assetnumber & VoD asset encoding version \\
\hline
hit & HTTP request was a cache hit or miss \\
\hline
cachename & cache's hostname \\
\hline
popname & cache's location \\
\hline
uid & id unique to a single user \\
\hline

\end{tabular}
\caption{List of log line features}
\label{tab1}

\end{table} 

\subsection{Data pre-processing}
%some equations in this section...

\subsubsection{Explicit vs. implicit feedback}

Rating about a product or service on a given scale (e.g., 1 to 5) is often a popular kind of explicit feedback. However, users are usually not easily convinced to give explicit feedback. Other kind of feedback (e.g., views, number of clicks, purchases) is often recorded by systems \cite{Rendle2012BPR_OPT}. In our case, we can extract explicit feedback from the CDN logs pertaining to number of requests for certain content. The cache logs contain a user identifier and the content requested, e.g., live TV channel or video on demand content identifier. This information was used to derive a user-item interaction feature to facilitate our matrix factorization modeling for ratings prediction. Values for the feature were on a log-scale of total requests.

As recommender systems models operate with (user, item) interaction samples, we performed the following mappings from the CDN log data:

\[ uid \Longrightarrow user \]
\[ livechannel/contentpackage \Longrightarrow item \]
\[ access frequency \Longrightarrow interaction \]

We grouped user and item Ids and established frequencies for each pair, resulting into a requests feature. We then put the request feature into a log-scale to get the interaction (or preference) feature. We used these features to create a user-item interaction matrix (similar to a utility matrix in recommender systems) as an input to our custom matrix factorization algorithm.
Inputs to matrix factorization algorithms of MyMediaLite platform are comma separated value (CSV) files consisting of the userId, itemId and interaction features without column or index names.
The results of this approach are depicted in Table \ref{tab2} showing user, item, frequency and engineered interaction values.

\begin{table}[h]
    \centering
    \begin{tabular}{|m{1cm}|m{2cm}|m{2cm}|m{2cm}|m{2cm}|}
    \hline
        No & userId & itemId & requests & interaction \\
        \hline\hline
        0 & 0 & 0 & 16634 & 10\\
        \hline
        1 & 1 & 1 & 18038 & 10\\
        \hline
        2 & 2 & 2 & 3019 & 8\\
        \hline
        ... & ... & ... & ... & ...\\
        \hline
        1978 & 1068 & 2 & 8 & 2\\
        \hline
        1979 & 1069 & 13 & 7 & 2\\
        \hline
    \end{tabular}
    \caption{Dataset view for user and item with corresponding interaction}\label{tab2}
\end{table}

% In the input dataset we identified 397 unique users and 74 unique items. Thus the shape of the utility matrix was $397 \times 74$ (397 rows of users, to 74 columns of items).
% Remember to use this in the Experiments Section

\subsubsection{Splitting Dataset into Train and Test Sets}
A random split of the datasets was performed on 70\%-30\% ratio for training and testing sets respectively (temporal trends were not considered).

%add the number/shape of the dataset.

\subsection{Baseline Methods}

Matrix factorization is the task of approximating a matrix $X$ by the product of two smaller matrices $W$ and $H$, that is $X \approx WH^T$. In the context of recommender systems, the matrix $X$ is partially observed ratings matrix, $W \in \mathbb{R}^{U\times K}$ where each row $U$ is a vector of users with $K$ latent factors, and $H \in \mathbb{R}^{I\times K}$ describes row $I$ of items, again with $K$ latent factors \cite{Nguyen2011MatrFact}. Thus, by finding the best $K$ latent features, one should be able to predict a rating with respect to a certain user and item, since the number of features associated with the users match with the number of features associated with the items \cite{Takacs2008Matr_Fact}. Let $w_{uk}$ and $h_{ik}$ be elements of $W$ and $H$, thus the rating of user $u$ to item $i$ is predicted by:

\begin{equation} \hat{r}_{ui}=\sum_{k=1}^{K}w_{uk}h_{ik}=(WH^T)_{u,i}  \end{equation}

where $W$ and $H$ are the model parameters, $\hat{r}_{ui}$ is the predicted rating, $r_{ui}$ is the actual rating from Table \ref{tab2} interaction column of user $u$ to item $i$. Model parameters can be learned by optimizing the objective function given a criterion such as root mean square error (RMSE) on a set of known ratings, thus reaching the goal of generalizing those ratings in a way that predicts future, unknown ratings/interactions:

\begin{equation} \min_{W,H}(r_{ui}-\hat{r}_{ui})^2 + \beta(\|W\|^2 + \|H\|^2)  \end{equation}

where $\beta$ is a regularization term to mitigate overfitting. For this experiment stochastic gradient descent was used in the optimization process for RMSE.

Initially, we implemented our own matrix factorization class. MyMediaLite recommender system platform's basic matrix factorization algorithm was also utilized. MyMediaLite is free and open-source software which consists of various state-of-the-art recommender system algorithms (e.g., matrix factorization, k-nearest neighbour, most popular item) and an extensive evaluation framework. It runs on the .NET platform, is written in the C\# (C-sharp) programming language \cite{Rendle2012BPR_OPT}.
MyMediaLite addresses two scenarios in collaborative filtering, rating prediction and item prediction from positive-only feedback. It offers state-of-the-art algorithms, online updates to already trained models, serialization of computed models and various routines for evaluation \cite{Gantner2011MyMediaLite}.
Beside basic matrix factorization model described earlier, we also utilized the biased matrix factorization technique to factor in user and item biases \cite{Koren2009MFTech}. The updated model then becomes:

\begin{equation} \hat{r}_{ui}=\mu + b_u + b_i + \sum_{k=1}^{K}w_{uk}h_{ik} \end{equation}

where $\mu$, $b_u$ and $b_i$ are global average, user and item biases respectively. We implemented our experiments in the Python programming language.

\subsection{Evaluation Techniques}

\begin{comment}
For model evaluation, metrics such as RMSE, Mean Average Error (MAE) (in case of rating prediction) and Receiver Operating Characteristics-Area Under a Curve (ROC-AUC), precision-at-N (prec@N), Mean Average Precision (MAP) and Normalized Discounted Cumulative Gain (NDCG).
\end{comment}

We evaluated our matrix factorization model with RMSE measure. For this metric, a low score is preferred. In this work, model parameters were optimized for RMSE calculation using stochastic gradient descent, based on the following equation:

\begin{equation} RMSE=\sqrt{\frac{\sum_{{ui}\in D^{test}}( \,r_{ui} - \hat{r}_{ui})\,}{|D^{test}|}} \end{equation}

The selection of RMSE metric was inspired by the 2009 Netflix competition which was won by a team of researchers, called “Bellkor’s Pragmatic Chaos” (10\% improvement over CineMatch RMSE) \cite{Takacs2008Matr_Fact}.

Evaluation protocols (splitting, candidate selection, metrics) are not easy to get right sometimes. The use of a random seed ensures that the comparison is done on the same “random” split. It can help in aspects such as debugging. We reuse methods/techniques to ensure comparability (more configurations kept fixed thus, less risk of accidental differences)\cite{Gantner2011MyMediaLite}. For this experiment a random seed of {$42$} was used and GitHub\footnote{\url{https://github.com/CDNResearchProject/Cache_Optimization/blob/main/Notebooks/Rating_Predict_LiveTV.ipynb}} was used as a version control system (scripts and configurations).

\subsection{Hyper-parameter Tuning}

Similarly to other machine learning techniques, the optimal selection of hyper-parameters are very important for matrix factorization. Hyper-parameters for matrix factorization models are latent factors ($K$), the learning rate ($\alpha$), a constant whose value determines the rate of approaching the minimum during parameter search, and regularization ($\beta$), to constrain the model from overfitting. The number of latent features ($K$) is often lower than the number of users and items. The optimal hyper-parameters were found through a grid search. The grid search method attempts all possible combinations of the provided range of hyper-parameter values \cite{Hsu2016SVM_Guide}.

\section{Results}

\subsection{Live-TV Content Analysis}

In the initial set of experiments, explicit feedback representation of the data was engineered for the interactions by using request frequencies for CDN content as described in Section 3.2, with Table \ref{tab2} providing a snapshot of the dataset. The dataset utilized initially pertained to liveTV content.

For this experiment, 5,261,828 cache log lines were used. After grouping uid with Live-TV content Ids there were 426,356 rows. This consisted of unique 137,654 user ids and 116 unique Live-TV ids. As described in Section 3.2, the split of train set and test set was on a 70\% to 30\% ratio respectively. The actual number of records is shown in Table \ref{tab3}.

\begin{table}[h!]
    \centering
    \begin{tabular}{|m{2.5cm}|m{2cm}|m{2cm}|}
    \hline
        LiveTV Datasets & \#Attributes & \#Records \\
        \hline\hline
        Train & 4 & 298,449 \\
        \hline
        Test & 4 & 127,907 \\
        \hline
    \end{tabular}
    \caption{Live TV datasets train-test-split}\label{tab3}
\end{table}

A custom Matrix Factorization class was written to perform matrix factorization. The main MyMediaLite library ('MyMediaLite.dll') was imported and used in Python.

\begin{figure}[h!]
\centering
\includegraphics[scale=0.5]{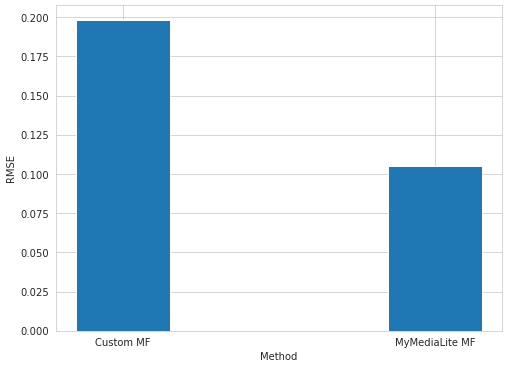}
\caption{RMSE measures for basic implemented matrix factorization and MyMediaLite matrix factorization on LiveTV content} \label{fig2}
\end{figure}

Table \ref{tab5} presents the comparison of RMSE of custom implemented matrix factorization and MyMediaLite matrix factorization during preliminary experiments. The results are visualized in the Figure \ref{fig2} as well.

\begin{table}[h!]
    \centering
    \begin{tabular}{|m{.5cm}|m{5cm}|m{3cm}|}
    \hline
        No & Method & RMSE \\
        \hline\hline
        1 & Custom matrix factorization class & 0.1981194 \\
        \hline
        2 & MyMediaLite matrix factorization & 0.1048596 \\
        \hline
    \end{tabular}
    \caption{RMSE measures for initial experiments on LiveTV content}\label{tab5}
\end{table}

\subsection{Live-TV and VoD Analysis}

We performed further experiments with both LiveTV (described in Section 4.1) and video on demand (VoD) content following the same design for initial set of experiments. 3,984,430 cache log lines for VoD were used. After grouping uid with VoD content Ids there were 206,334 rows. This consisted of unique 187,302 user ids and 3940 unique VoD ids. Table \ref{tab4} shows the actual number of records after the random split

\begin{table}[h!]
    \centering
    \begin{tabular}{|m{2.5cm}|m{2cm}|m{2cm}|}
    \hline
        VoD Datasets & \#Attributes & \#Records \\
        \hline\hline
        Train & 4 & 144,433 \\
        \hline
        Test & 4 & 61,901 \\
        \hline
    \end{tabular}
    \caption{VoD datasets train-test-split}\label{tab4}
\end{table}

%\subsection{Baseline Methods} 
%weighted regularized MF, bayesian personalized ranking MF, non-negative MF

\begin{figure}[h!]
\centering
\includegraphics[scale=0.5]{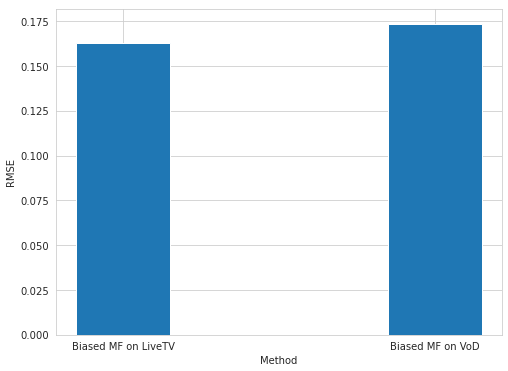}
\caption{RMSE measures of biased matrix factorization on LiveTV and VoD content} \label{fig3}
\end{figure}

Results for biased matrix factorization experiments with LiveTV and VoD datasets are provided in Table \ref{tab6} and corresponding visualization on Figure \ref{fig3}

\begin{table}[h!]
    \centering
    \begin{tabular}{|m{.5cm}|m{5cm}|m{3cm}|}
    \hline
        No & Method & RMSE \\
        \hline\hline
        1 & Biased MF (LiveTV) & 0.1628978 \\
        \hline
        2 & Biased MF (VoD) & 0.1734011 \\
        \hline
    \end{tabular}
    \caption{RMSE measures for further experiments on LiveTV and VoD content}\label{tab6}
\end{table}

We used a custom grid search code compatible with our matrix factorization implementation to perform brute-force search for the best hyper-parameters. MyMediaLite platform contains its own class for grid search which aided in locating the best hyper-parameters.
For referencing, the best hyper-parameters that were found in the cross-validation for Experiment 1 are shown in Table \ref{tab7} and in Experiment 2 in Table \ref{tab8}.

\begin{table}[h!]
    \centering
    \begin{tabular}{|m{.5cm}|m{5cm}|m{5.5cm}|}
    \hline
        No & Method & Hyper-parameters \\
        \hline\hline
        1 & Custom class matrix factorization & $K=46$, $\alpha=0.3$, $\beta=0.02$, $iter=100$ \\
        \hline
        2 & MyMediaLite matrix factorization & $K=51$, $\alpha=0.3$, $\beta=0.01$, $iter=100$ \\
        \hline
    \end{tabular}
    \caption{Hyper-parameters for LiveTV preliminary experiments}\label{tab7}
\end{table}

\begin{table}[h!]
    \centering
    \begin{tabular}{|m{.5cm}|m{5cm}|m{5.5cm}|}
    \hline
        No & Method & Hyper-parameters \\
        \hline\hline
        1 & Biased MF on LiveTV & $K=52$, $\alpha=0.07$, $\beta=0.05$, $iter=100$ \\
        \hline
        2 & Biased MF on VoD & $K=60$, $\alpha=0.07$, $\beta=0.06$, $iter=100$ \\
        \hline
    \end{tabular}
    \caption{Hyper-parameters for LiveTV and VoD content: MyMediaLite platform}\label{tab8}
\end{table}

In results of Experiment 1 for Live-TV content analysis, the RMSE measure using algorithm in MyMediaLite is better than our custom implementation of the basic matrix factorization algorithm as shown in Table \ref{tab4} and Figure \ref{fig2}. In Experiment 2 for Live-TV and VoD content analysis, with the biased matrix factorization algorithm of MyMediaLite platform, RMSE measures are comparable in the case of LiveTV and VoD datasets, as shown in Table \ref{tab6} and visualized in Figure \ref{fig3}. These RMSE scores imply good precision in predicting rating for cache content, thus cost reduction potential in real-life content delivery scenarios. With these matrix factorization models, CDN operators will be equipped with the enhanced capability to predict which live-TV and VoD contents are popular based on ratings (interaction values) and thus to be cached.

\section{Conclusion}

The goal of this research was to show that recommender engines, and matrix factorization in specific, can be utilized in the context of predicting popular content and potentially optimizing cache admission/eviction in a content delivery network setting. We extracted user and content identifiers from CDN logs and generated ratings based on the frequency of requests targeting specific resources (e.g., live TV channels or video on demand content). We used that data to create a matrix factorization model. We showed that a low root mean square error value can be achieved on real-life CDN log data. 
Future prospects for this work are to leverage CDN logs to further preprocessing, and engineering the ratings in specific, and develop models for item prediction from positive-only implicit feedback which we believe would complement the results presented in this paper.

%
% ---- Bibliography ----
%
%\begin{thebibliography}{5}
%

%\printbibliography
\bibliography{CDNreferences}
\bibliographystyle{splncs04}

\end{document}